# Vito Volterra and his commemoration for the centenary of Faraday's discovery of electromagnetic induction

Amelia Carolina Sparavigna
Politecnico di Torino

**Abstract:** The paper presents a memoir of 1931 written by Vito Volterra on the Italian physicists of the nineteenth century and the researches these scientists made after the discoveries of Michael Faraday on electromagnetism. Here, the memoir entitled "I fisici italiani e le ricerche di Faraday" is translated from Italian. It was written to commemorate the centenary of Faraday's discovery of the electromagnetic induction. Besides being a remarkable article on the history of science, it was also, in a certain extent, a political paper. In fact, in 1931, the same year of the publication of this article, Mussolini imposed a mandatory oath of loyalty to Italian academies. Volterra was one of the very few professors who refused to take this oath of loyalty. Because of the political situation in Italy, Volterra wanted to end his paper sending a message to the scientists of the world, telling that the feeling of admiration and gratitude that in Italy the scientists had towards "the great thinker and British experimentalist" was profound and unanimous.

**Introduction**
"The discoveries made by Faraday were so numerous and important that they originated new and important developments of physics and inspired the most varied and admirable practical applications. The concepts that guided him in his famous experiments deeply changed the way we think of natural phenomena, so that a new era in the history of natural philosophy begins with him". These are the opening phrases that we can find in a memoir written by Vito Volterra, entitled "I fisici italiani e le ricerche di Faraday". The subject of the memoir was the physics of electromagnetism stimulated in Italy by the discoveries of Michael Faraday [1]. This paper, written in 1931 to commemorate the centenary of Faraday's discoveries, shows that Volterra had a deep knowledge of the researches on electromagnetism made by Italian physicists in the nineteenth century.
Vito Volterra (1860 – 1940) was an Italian mathematician and physicist. Born in Ancona, Volterra studied at the University of Pisa, where he became professor of rational mechanics in 1883. In 1892, he became professor at the University of Torino and then, in 1900, professor at the University of Rome La Sapienza. He started his researches with the study of functionals, after he worked on integral and integro-differential equations. He reported his researches in Theory of functionals and of Integral and Integro-Differential Equations (1930) [2]. In 1905, he started studying the dislocations in crystals, a research that is fundamental for understanding the behavior of ductile materials. During the World War I, he worked for the Italian Army investigating the use of inert helium rather than flammable hydrogen in airships. After the war, he turned his attention to applying mathematics to biology, obtaining the Lotka–Volterra equations. In 1922, he joined the opposition to the Fascist regime of Benito Mussolini.
In 1931 he was one of only 12 out of 1,250 professors who refused to take a mandatory oath of loyalty [3]. And it was in this year, that he published the paper [1], the translation of which we are here proposing. At a first glance, this paper of Volterra is a very interesting memoir on the history of physics in Italy. However, it had also a political aim. Because of the political situation in Italy, Volterra wanted to send a message to the scientists of the world, ending his memoir on the results that the Italian scientists obtained as a consequence of the Faraday's discoveries, telling that the "feeling of admiration and gratitude that in Italy we have towards the great thinker and British experimentalist are profound and unanimous".

Let us start reading the Volterra's article, of which I give a new translation. A translation of it was published in August 1931 by Nature (doi:10.1038/128342a0), under the title "Italian Physicists and Faraday's Researches" (it is not freely available on the web). Let me stress that I have not seen this translation nor used it for my translation. I used only the Italian text provided by http://www.liberliber.it/online/autori/autori-v/vito-volterra/i-fisici-italiani-e-le-ricerche-di-faraday/

The references the reader find in the following text are those given by Volterra.

**Volterra's article**

The discoveries made by Faraday were so numerous and important that they originated new and important developments of physics and inspired the most varied and admirable practical applications. The concepts that guided him in his famous experiments deeply changed the way we think of natural phenomena, so that a new era in the history of natural philosophy begins with him.

In the recent years, any scholar of physics had been, more or less, involved in Faraday's works. So, if we would mention all those who were followers of his work and that have made use of his results, we should make the name of all the physicists of recent times. This is true for Italy such as for other countries. In fact, it is not the case of reporting the history of Physics in Italy in the last century, but here I want to give report of the investigations that were more directly inspired by those of Faraday, especially during their first announcement, and had great successful and important applications.

I will refer therefore to August 1831, when the attempts of Faraday to have currents through the action of magnets had finally a successful outcome. The result was not communicated to the Royal Society of London until the November of the same year, but the memoir was published later and the French translation appeared only in May 1832 in the "Annales de Chimie et de Physique". Faraday, to repair to this delay, wrote a very concise letter on this subject to the Hachette of Paris, that made it known to the Accademia delle Scienze in December 1831.

Nobili, a physicist of considerable value of the Museum of Florence, already known, among other things, for his astatic galvanometer, his rings and studies on thermo-electricity, heard of this discovery from Amici, who read it in "Temps". Understanding immediately the importance of this discovery, Nobili began to repeat the Faraday's experiences, joining in this work another Florentine physicist, Antinori. Their first memoir dated January 1832, and it brings explicitly this date, but it was published in the November 1831 issue "Antologia", which came out very later [4]. The following memories appeared in the same newspaper.

These two physicists of Florence obtained induced currents, either approaching a closed circuit to the pole of a magnet, or by opening a circuit constituted by a horseshoe magnet and the support of this magnet, or by rotating upside-down a coil circuit arranged parallel to a magnetic inclination needle, or by introducing a soft iron core in a coil which was forming a closed circuit. These experiences aroused a controversy (one of the few controversies of Faraday), because of the considerations that accompanied them, especially those referring to the Arago's magnetism of rotation and to the spark of induction. The tardy knowledge of the specific text of the Faraday's Memoir, and the incorrect interpretation of the letter to Hachette, influenced in creating those misunderstandings which originated, as noted by Naccari, the mentioned controversy [5].

The studies on induction continued in Italy especially with the works of Matteucci, who investigated the distribution of the currents in the rotating disk of Arago and the magnetism of rotation. In 1854 he published his "Cours spécial sur l'induction, le magnétisme de rotation" that was reporting what he knew at the time on the phenomena of electromagnetic induction [6].

However, a work existed that had a large philosophical and experimental importance and of this work I wish to renew the memory. In 1852, Felici, at the time assistant of Matteucci, and after successor, in the Department of Physics at the University of Pisa, began the publication of three memoirs entitled: Sulla teoria matematica dell'induzione elettro-dinamica [7]. F. Neumann had already published in 1845-47, his famous formula in two works, and in 1846 it appeared that of Weber. Felici abandoned the ways taken by these scientists and set out to get the formulas giving the laws of electromagnetic induction, following, step by step, the purely experimental method. This was the way that, a quarter of a century before, Ampere used to establish the formula of the ponderomotive force existing carrying between them two currents.

Like Ampere, in his first memoir Felici used solely equilibrium experiments. With an ingenious method, he enhances the inductive action, using repetitive open/closed current, adding in the galvanometer the intensities of the repeated induced currents having the same flowing direction. In this manner, he obtained a formula that gives the electromotive force corresponding to the action of the elements of current, formula which contains a contribution which is null when the electromotive force is integrated around closed circuits. In this manner, the Felici's formula is confirming experimentally the formula that Neumann obtained extending the Lenz's law and tell us something more.

In his following Memories, Felici studies the induced currents, not only in filiform circuits, but also in conductors having more than one dimension, linking these studies to the experimental researches of Matteucci exposed in his "Cours special" that we have above mentioned.

In the same period, during which Matteucci and Felici flourished in Pisa, Ottaviano Fabrizio Mossotti was also teaching there. He was one of the most prominent Italian personalities of the last century. Born in Novara in 1791, he graduated in Pavia and there he studied with Brunacci and also attended the lessons of Volta. Abandoned Italy as a result of political persecutions, he went to France and England, where he was received with great respect and had many friends. Called first in Argentina, then he became professor in Corfu, in the period of the English protection, and finally he arrived in Pisa, where he founded a school of astronomy, physics and mathematics, which was of great importance in the recent history of science in Italy.

Among the several beautiful researches of Mossotti, today, the one which seems the most important and that gives him fame is the theory of dielectrics. Faraday devoted the eleventh and the twelfth series of his "Experimental Researches on Electricity" to study the role of dielectrics in the electrostatic induction. These features of dielectrics had already called the attention of Avogadro for several years [8], but the famous and decisive Faraday's experiments showed their nature and led to the theory of the polarization of dielectrics.

We must credit to Mossotti that the theory of electrical induction in dielectrics was linked to the theory of the magnetic induction. Let us remember that Poisson had given the mathematical theory of the magnetic flux. Therefore, it was enough to read in the language of electricity what Poisson had obtained in terms of magnetism, to have a mathematical theory for dielectrics based on the Faraday's concepts. Then, a part of the Mossotti's memoir of 1846 [9] is, according to the same author, a repetition of that of Poisson. The remarkable and elegant consequences that Mossotti obtained and the interesting applications of other researchers, for example that of Clausius on the back discharges, make the Mossotti's work possessing a specific value.

Such work is linked to the Mossotti's concepts on the constitution of matter, subject which this author has considered several times, both in a memoir dedicated to Plana [10], both in the opening address at the Corfu lessons [11], and in other occasions. Faraday too, in the nineteenth series of his Experimental Researches, welcomes these Mossotti's ideas that are in some extent corresponding to his owns. These ideas deserve to be read again today, and regarded and compared with the latest theories. The fact that they had been devised before the development of thermodynamics make them distant from today's conception of many phenomena; however, in spite of this, they may contain some seeds for fruitful thoughts.

The studies on dielectrics, so cleverly explored through the mathematical analysis by Mossotti, continued in Italy with some works, of which I mention those of Belli, Matteucci, and Felici on the dielectric currents, and Felici on the viscosity of dielectrics. Such researches have continued till the present day, with the works, both theoretical and practical, of physicists and electro-technical researchers.

The analysis of Mossotti is an example of the potentiality the Faraday's ideas have to be carried out in the mathematical formalism. This is not the only example, nor the most famous, and in fact everybody knows that Maxwell was able of translating the concepts of Faraday in the equations of the electromagnetic field, and found in them the germ of the electromagnetic theory of light.

Faraday did not know the techniques of geometry, but he knew exactly how to put into words what the formulas can say. The same can be repeated for Volta. If, in the manner of Plutarch, we wanted to make a parallel between these two heroes of science, it could be possible to stress the similarity, in this regard, of the minds of the British Physicist and of the Italian Physicist. In fact, they were so similar in other ways too. Their minds were both mathematical, in a deeper and more intimate sense, since symbols and analytical artifices are not the substance, but only the outward appearance of mathematics.

The problem of the electric motor and the inverse problem, that is, that of producing electric currents, first generated by Volta's piles, making use of the Faraday induction principle with a method capable of being applied on a large scale, were topical problems for sure, when an ingenious solution was given in 1860 by Antonio Pacinotti, who built his famous ring in the laboratory of Technological Physics at the University of Pisa. Only in 1864, the discoverer put down, in a written text, the description of the apparatus, which he modestly called "an electromagnetic machine" and published in 1865 in the "Nuovo Cimento" [12].

I feel no need of mentioning here the description of this machine, because it is reported in every elementary treatise of Physics and Electrical Engineering. The brilliant solution of the problem proved itself useful for wider and important industrial applications, immediately perceived by the inventor himself, and it became one of the most useful machines and largely used in the world. There is no need to repeat the well-known story of its rapid spread and controversies, now ended, which it gave rise.

As we have told before, the researches concerning the theories of Maxwell cannot be separated from the memory of Faraday, being the theories themselves derived from his experiences and concepts. The light pressure, that Maxwell obtained as a result of electromagnetic theory, was demonstrated with a new and original method by Bartoli in a Memoir, published in 1876, devoted, in a special manner, to the study of the radiometer [13]. Bartoli imagines four spherical concentric shells: the outermost and the innermost perfectly black, the two intermediate shells perfectly reflecting. Through a cycle which can be repeated as often as we desire, he can pass heat from the outer shell, which can be supposed as the colder, to the innermost, which can be supposed as the hotter. The second principle of thermodynamics requires that, for a cycle, a quantity of work must be converted into heat and that then, being the inner body irradiated, a pressure is necessary being exerted on the adjacent shell. From this fact, it follows as a consequence the pressure of light. Of the several works of Bartoli, this is the most famous and his name is undoubtedly linked to it. It is remarkable, however, that the Bartoli, who died in July 1896, before the pressure of light was experimentally verified, was doubting its reality and he tried to interpret in different manner his arguments, but, during his life, he always concealed these doubts that became apparent only later.

We cannot ignore the beautiful researches of Bartoli on electrolysis and in particular those on the ability of decomposing water even with weaker electromotive forces [14]. They are connected to the studies of Faraday of the fifth series and subsequent series.

Original feature of Faraday's thought was that of transporting the seat of electromagnetic phenomena in the dielectric material, where he saw and imagined the evolution and changes of the lines of force, which were materializing what he conceived to explain the electrodynamic actions.

One of the brightest and most effective exposure of the Faraday's concepts, in the manner by which they were performed by Maxwell and Hertz, was the speech given by Galileo Ferraris at the Accademia dei Lincei in June 1894 on the transmission of electricity [15]. However, such a speech would be just the memory of a happy popularization of the concepts of localization and flow of energy, that at the time were new concepts for a large public audience, if it did not serve to reveal much more, that is, the intimate manner by which Ferraris understood the nature of electrical phenomena, the starting point of the discovery of the rotating magnetic field which he made nine years earlier. And in fact, it was the vision by mind's eyes of the electrical oscillations, not inside the conductors but in the space out of them, and the consequent persuasion of the possibility of combining two orthogonal electrical oscillations like combining those of light - a vision that appeared as a flash while he was walking alone at night down the streets of Torino - which gave rise to the discovery to which Ferraris mainly owes his reputation [16].

This brilliant achievement was preceded by long and deep studies of Ferraris on the transformers, of which he gave a complete theory that led him to establish the law of the phase shift and to calculate the power of a transformer. This was enough to dissipate the erroneous opinions that existed on the performance of a transformer and to show the importance in industrial applications of this fundamental device, derived directly from the induction principle, which, in electrical engineering, is playing the role of the lever in mechanics [17]. The transport of energy with the use of alternating currents, from which all industries received a large impulse and benefit, was the practical result of these studies.

The history of the evolution of Faraday's ideas, which led to the mathematical theory of Maxwell, and then to the experiences of Hertz, would not be complete if we do not talk of the wireless telegraph that is so wonderfully closing such memorable cycle. In Italy, Righi, among others, repeated and continued successfully the experiments of Hertz, pursuing, in all details, the similarities between the optical phenomena and those of electromagnetic waves [18]. Marconi then, with admirable perseverance, and taking advantage of brilliant and various procedures, managed to achieve practical effects of the remote transmission methods using electromagnetic waves. Rapid advances in wireless telegraphy and radiotelephony are held daily before our eyes, and are followed with interest and wonder in all the world.

I cannot continue on the discussion of several other issues, although of great interest, such as electrolysis and electrophysiology, evoking the studies of Matteucci [19] and the correspondence with Faraday [20], or mentioning the chemical theory of the pile that Faraday preferred to that of the contact, linking myself to the ideas first expressed by Fabbroni [21]. In the same manner, I cannot talk about the issue, debated by Faraday, if the electricity, however produced, is always the same and follows the same laws, giving equal effects, of which the same Volta discussed, and that gave rise among us to recent studies too. I am just mentioning the observations of Father Bancalari in 1847 [22] and the subsequent memoir of Zantedeschi on the magnetism of flames [23], works that attracted the interest of Faraday on magnetism and diamagnetism of gases.

Finally, I am closing the paper by mentioning the magneto-optics, which starts from one of the most famous discoveries of Faraday, that of the magnetic rotation of polarization. Righi gave a kinematic explanation of the phenomenon, that further researches linked to the great discovery made by Zeeman. And Righi himself made use of the sodium flame, crossed in the direction of the field by a beam of light between two crossed Nicol filters, a flame which allows, with simple devices, the observation of the Zeeman inverse phenomenon [24].

It is very remarkable that this experience has been carried out with a device similar to one devised by Faraday, who, however, escaped the final result, a very rare episode indeed during the researches of this great observer. I'll stop here, after quoting the late persons that in Italy took care of this subject, and pointing out that other scientists of value exist that gave and are giving valuable contributions. Like a few years ago, the work of Volta and that of Ampere, so today that of Faraday is recalled and commemorated.

These three great scientists, from different countries, lived far away and separated from each other, but their thoughts concurred harmoniously to the advancement of natural philosophy and the discovery of new and wonderful findings useful to human society. The influence that each one exercised beyond the borders of their homeland served to drive towards the similar goals the mind of different people, who gained new worth in the common effort.

The aim of this brief essay was that of highlighting the influence that the work of Faraday had in Italy. From a philosophical point of view, it was very great and, as we have seen above, the discoveries of the English physicist penetrated among us and aroused new and original researches, of singular importance from the point of view of practical applications, which lifted the enthusiasm not only in men of science, but also in the whole society, that from them gained many benefits. The feelings of admiration and gratitude that in Italy we have towards the great thinker and British experimentalist are profound and unanimous.

**References**
[1] Vito Volterra, I fisici italiani e le ricerche di Faraday, «L'Elettrotecnica», vol. XVIII, 1931, pp. 806–808.
[2] Vv. Aa. (2016). Wikipedia. URL: https://en.wikipedia.org/wiki/Vito_Volterra
[3] Simonetta Fiori, I professori che dissero no a Mussolini, la Repubblica.it , 14 Aprile 2000.
[4] Sopra la forza elettromotrice dal magnetismo dei sigg. L. Nobili e V. Antinori, «Antologia», n. CXXXI; Nuove esperienze elettromagnetiche e teoria fisica del magnetismo di rotazione, «Antologia», n. CXXXIV; Descrizione delle nuove calamite elettriche ed osservazioni sulle medesime, «Antologia», n. CXXXVI; Sopra varii punti di magneto elettricismo, «Antologia», n. CXXXVIII; Teoria fisica delle induzioni elettrodinamiche di L. Nobili, «Antologia», n. CXLII.
[5] La vita di Michele Faraday, Padova, 1908.
[6] Cours spécial sur l'induction, le magnétisme de rotation, le diamagnétisme et sur les relations entre la force magnétique et les actions moléculaires, Paris, 1854.
[7] «Annali delle Università Toscane», vol. 3°. Le tre Memorie del Felici furono tradotte in tedesco e stampate negli «Ostwald Klassiker der exakten Wissenschaften».
[8] Considérations sur l'état dans lequel doit se trover une couche d'un corps non conducteur de l'électricité lorsqu'elle est interposée entre deux surfaces douées d'électricité de différente espèce, «Journal de Physique, de Chimie et d'Histoire Naturelle et des Arts», par J. C. Delamétherie, Paris, vol. 63, Juillet, 1806; Suite des considérations, etc.; Avogadro, Saggio di teoria matematica della distribuzione dell'elettricità sulla superficie dei corpi conduttori nell'ipotesi dell'azione induttiva esercitata dalla medesima sui corpi circostanti per mezzo delle particelle dell'aria frapposta, «Memorie Società Italiana delle Scienze», vol. 23, 1844.
[9] Discussione analitica dell'influenza che l'azione di un mezzo dielettrico ha sulla distribuzione dell'elettricità alla superficie di più corpi elettrici disseminati in esso, «Memorie della Società Italiana delle Scienze», Modena, Parte 2a, vol. 24, 1846.
[10] Sur les forces qui régissent la constitution intérieure des corps, aperçu pour servir à la détermination de la cause et des lois de l'action moléculare, Taylor's, Torino, 1836.
[11] Lezioni elementari di fisica matematica, Firenze, 1843.
[12] Descrizione di una macchinetta elettromagnetica del dott. Antonio Pacinotti, «Nuovo Cimento», giugno 1864, pubblicato il 3 maggio 1865.
[13] Sopra i movimenti prodotti dalla luce e dal calore e sopra il radiometro di Crookes, Firenze 1876.
[14] Citiamo fra le varie Memorie del Bartoli su questo soggetto: Su le polarità galvaniche e su la decomposizione dell'acqua con una pila di forza elettromotrice inferiore a quella di un elemento Daniell, «Nuovo Cimento», 1879.

[15] Lettura fatta alla R. Accademia dei Lincei nella solenne adunanza del 3 giugno 1894.

[16] Rotazioni elettrodinamiche indotte per mezzo di correnti alternate, «Atti della R. Accademia delle Scienze di Torino», vol. 23, 18 marzo 1888.

[17] Ricerche teoriche e sperimentali sul generatore secondario Gaulard e Gibbs, «Memorie della R. Accademia delle Scienze di Torino», vol. 37, serie 2, genn. 11, 1885; Sulle differenze di fase delle correnti, sul ritardo dell'induzione e sulla dissipazione di energia nei trasformatori, «Memorie della R. Accademia delle Scienze di Torino», dic. 4, 1887.

[18] L'ottica delle oscillazioni elettriche, Bologna, 1887.

[19] Electro-physiological Researches, «Phil. Trans.», 1845-1830; Lezioni sui fenomeni fisico-chimici dei corpi viventi, Pisa, 1846.

[20] Dott. Bence Jones: The Life and letters of Faraday, vol. 2°.

[21] Sur l'action chimique des differents métaux à la temperature de l'atmosphère et sur l'explication de quelques phénomènes galvaniques, Paris, 1796; Dell'azione chimica dei metalli nuovamente avvertita, Firenze, 1793.

[22] Ueber eine Entdeckung des Magnetismus der Flamme, «Pogg. Ann.», vol. 73.

[23] «Raccolta fisico-chimica Italiana», vol. 3°. Circa le esperienze dello Zantedeschi che preluderebbero all'induzione elettrodinamica, cfr. «Biblioteca Italiana», vol. 53 e Naccari, op. cit., pag. 236.

[24] Di un nuovo metodo sperimentale per lo studio dell'assorbimento della luce nel campo magnetico; 2 Note, «Rend. Acc. dei Lincei», serie V, vol. 7°, 2° sem., 1898.